# Enhance the machine learning algorithm performance in phishing detection with keyword features

Zijiang YANG
*Department of Computer Science*
*New York University*
New York, United States
zy3110@nyu.edu

*Abstract*—Recently, we can observe a significant increase of the phishing attacks in the Internet. In a typical phishing attack, the attacker sets up a malicious website that looks similar to the legitimate website in order to obtain the end-users' information. This may cause the leakage of the sensitive information and the financial loss for the end-users. To avoid such attacks, the early detection of these websites' URLs is vital and necessary. Previous researchers have proposed many machine learning algorithms to distinguish the phishing URLs from the legitimate ones. In this paper, we would like to enhance these machine learning algorithms from the perspective of feature selection. We propose a novel method to incorporate the keyword features with the traditional features. This method is applied on multiple traditional machine learning algorithms and the experimental results have shown this method is useful and effective. On average, this method can reduce the classification error by 30% for the large dataset. Moreover, its enhancement is more significant for the small dataset. In addition, this method extracts the information from the URL and does not rely on the additional information provided by the third-part service. The best result for the machine learning algorithm using our proposed method has achieved the accuracy of 99.68%.

## I. Introduction

In the recent years, there is a significant increase of phishing attack in the Internet. Attackers use false websites that look like the legitimate websites to deceive the end-users so as to obtain the sensitive information, such as account names, passwords, security numbers and etc. According to the report published by Anti-Phishing Working Group(APWG), the number of observed phishing attacks is currently increasing exponentially[1]. Unfortunately, the financial sector continued to be the most-attacked sector and the financial loss caused by this kind of attack is estimated as much as billions of dollars annually. However, it is inevitable to prevent the end-users from clicking the URLs that direct to the false websites[1].

To protect the end-users from the phishing attacks, a number of methods for early detection of these malicious website URLs are developed in the literature. Previously, researchers use the blacklist approach to detect the phishing URLs[2]. Basically, the website visited by the end-user will be searched in the blacklist database. If the website is found, it will be classified as phishing. However, this approach requires frequent update of the database and the attackers can easily adjust the URL to bypass this detection mechanism[3].

Recently, the machine learning methods are proposed to address this problem[4][5][6]. Researchers use various approaches to extract features and to build up the machine learning classification algorithm to distinguish the phishing websites from the legitimate ones. The features extracted can be categorized into three categories: (1)URL-based: use the information of the website URL[7]; (2)Content-based: use the information of the webpage of the website, mainly the html, CSS and image information[8][9]; (3)Domain-based: use the information of the website domain, such as the response time of the website[10][11][12]. The classification algorithms are mainly the traditional ones and the ensemble of them. It is reported that the machine learning methods achieves high accuracy in this detection task.

This paper presents a novel hybrid approach that combines the traditional features with the keyword features to improve the accuracy of the existing machine learning algorithms. This approach has the following advantages: (1)use only the URL information; (2)improve the existing classification algorithms, especially in the small dataset classification tasks; (3)easy to extend to achieve better performance.

The rest of this paper is organized as follows: Section II introduces the related work about the phishing detection methods proposed by the previous researchers. Section III provides the details about the proposed method. Section IV presents the implementation and the analysis of the results. Finally, Section V concludes the work and discusses the future work.

## II. Related work

Researchers have proposed multiple methods in the past to solve the detection task of phishing website URL. These methods can be divided into two groups: (1) list-based detection systems; (2) machine learning based detection systems.

Cao et al. [13] developed a whitelist method that register the IP address of the websites. The end-users will be warned when visiting the website whose information is not consistent with the record. However, any legitimate website will be warned at the first time when the end-user visited. Sheng, S et al.[14] used the blacklist to detect the phishing URLs. Any website on the blacklist will be blocked. Apparently, this method cannot prevent the attacks from the websites that are newly set

up[15][16][17]. To obtain proper defense against these new websites, frequent update of the blacklist is required[5].

Another approach to solve this problem is to use the machine learning classification algorithms to distinguish the phishing URLs from the legitimate URLs. Fette et al[3] first introduced the machine learning algorithms to detect the phishing emails. Then, Abu-Nimeh et al[18] applied various machine learning algorithms to detecting the phishing websites. A. Le et al[11] extracted the features from the URLs to train the machine learning classifiers and achieve high accuracy. Rosiello et al[19] used the layout information of the websites to generate features for the phishing detection algorithms. Y. Zhang et al[20] specifically extracted the content from the websites to generate the features. Sami Smadi et al[21] and Ozgur Koray Sahingoz et al[22] modified the classical machine learning algorithms to enhance the classification accuracy. A. Basit et al[23] used the ensemble method to combine the opinions from various classifier so as to give better classification result.

Previous researchers have done a lot of work on modifying the existing machine learning algorithms. Not much research focuses on the feature generation. This paper will propose a novel hybrid approach that combines the traditional features with the keyword features to improve the accuracy of the existing machine learning algorithms.

## III. PROPOSED METHOD

### A. Design Objective

Existing machine learning phishing detection methods are mainly facing these challenges: (1)using third-party services(such as obtaining the age of the domain and the Google indexes); (2)needs to download the webpage information to analyze; (3)using computationally-heavy algorithms that are difficult to implement in real-time detection. On top of the previous research, we would like to develop a simple but effective phishing detection algorithm with machine learning. We want the method to achieve the following goals:

1. Builds up the algorithm using the URLs information only.
2. Extracts the effective features from the URLs by capturing the key word information.
3. Use the fast but also accurate detection algorithm

### B. Traditional Feature Extraction

For feature extraction, the first thing we do is to analyze the structure of a URL. As is shown in Fig.1, the URL can be decomposed into the following segments:

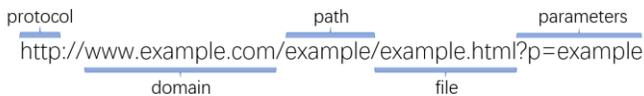

Fig. 1. Example of a URL.

Traditional approach counts the number of special character such as the dot(.), hyphen(-), slash(/) and percent(%) in the entire URL and each segment respectively. Also, the lengths of the domain, path, file and parameters are also included. Table 1 shows some selected features and their description.

TABLE I. SELECTED SAMPLE FEATURES FOR THE DATASET

| Feature Name | Description | Type |
| --- | --- | --- |
| url_dot | Number of "." signs in url | Numeric |
| url_hyphen | Number of "-" signs in url | Numeric |
| url_length | Number of characters in url | Numeric |
| domain_dot | Number of "." signs in domain | Numeric |
| domain_hyphen | Number of "-" signs in domain | Numeric |
| domain_length | Number of domain characters | Numeric |
| pathfile_dot | Number of "." signs in path and file | Numeric |
| pathfile_hyphen | Number of "-" signs in path and file | Numeric |
| pathfile_length | Number of path and file characters | Numeric |
| params_dot | Number of "." signs in parameters | Numeric |
| params_hyphen | Number of "-" signs in parameters | Numeric |
| params_length | Number of characters in parameters | Numeric |

### C. Key Word Feature Extraction

The algorithm we develop differs from the existing ones is to incorporate the key word features into the traditional features used in the literature.

*1) http keyword*

Count the number of key word 'http' in the URL. The phishing website cannot generate the website itself, it copies the content from the legitimate website. This characteristic can be captured in the URL. For example, the following is a phishing URL:

http://www.xmadwater.com.cn/js/?ref=http://us.battle.net/d3/en/index

this phishing website copies the content from us.battle.net website. Unlike the normal URL with only one 'http' keyword at the beginning, we can see that 'http' count would be 2 in this phishing URL.

*2) ref keyword*

Similarly, we capture the occurrence of the keyword 'ref' in the URL.

*3) login keyword*

We count the number of keyword 'login' in the URL as the phishing websites are interested in the end-users' login information.

*4) account keyword*

Similar to the 'login' key word, we count the number of key word 'account' in the URL, as the phishing website also wants the account information.

*5) apple keyword*

Nowadays, Apple ID is crucial to the Apple users, as it contains a lot of sensitive information, especially the Apple payment information. We count the keyword 'apple' in the URL.

*6) paypal keyword*

Similar to the 'apple' keyword, attackers want to obtain the payment information in the end-users' PayPal account. We count the keyword 'paypal' in the URL.

We append the above features to the traditional ones to enhance the machine learning algorithm performance. Apparently, the keyword features have their own contextual meaning in the URL, which is quite different from the traditional ones. In previous research, some researchers may use the Natural Language Processing(NLP) techniques to extract the words. However, the amount of the words is usually too large and they rely on some dimensional reduction mechanism to reduce the number of words. This process may take a long time to accomplish. Also, based on our experiments, adding too many keyword features will cause serious overfitting in the training sample. Keeping the keyword feature set concise, informative and meaningful is crucial to the enhancement of the algorithms.

The advantages of adding these keyword features are as follows: (1)use the URL information only; (2)capture the information that the traditional features don't contain; (3)easy to compute.

## IV. EXPERIMENTS AND RESULTS

### A. Dataset

We obtain the data from the public dataset which list out the URLs of the phishing and legitimate websites. This name of this dataset is ISCX-URL2016. It is from the University of New Brunskwick. The direct URL to data is:

https://www.unb.ca/cic/datasets/url-2016.html

The dataset consist of over 45,000 instances in which 35,378 instances are legitimate and the remaining 9,965 instances are phishing. To achieve a balanced dataset, we randomly extract out a subsample 10,000 legitimate URLs and combine it with the 9,965 phishing URLs. In the experiment, we randomly pick 80% of the data for training and 20% for testing. For each URL, we decompose it into the domain, path, file and parameters where the features are generated respectively. Each instance consists of 26 features, 20 of which are traditional features, while the remaining 6 features were extracted as the keyword feature.

### B. Machine Learning Algorithms

In the literature, various machine learning algorithms, basically the classification algorithms, are developed to detect the phishing URLs. In this paper, we list out the traditional algorithms are previously used.

*1) Random Forest*
Random forest uses bagging technique to aggregate a number of decision trees in parallel from the bootstrap samples of the data set. The final decision is the voting result of all the decision trees.

*2) Extreme Gradient Boosting Decision Trees(XGBoost)*
XGBoost also builds up a number of decision tree. It differs from the random forest in the way it builds the trees. It iteratively builds the tree so as to minimize the error between prediction result of the current forest and the target. The final decision is the weighted sum of all the decision trees.

*3) Multilayer Perceptron(MLP)*
It is an artificial neural network that mimic the structure and operation of the human brain. A number of layers of neurons including the input layer, the hidden layer and the output layer are constructed. The instance features are fed into the input layer. Signals are activated to the next layer through a non-linear activation function. The final classification results are given by the output layer. In this experiment, multilayer perceptron with 5 layers, in which 3 layers are the hidden layers, is constructed. The size of the layers for these 3 hidden layers is 40, 20 and 10. ReLU function is used as the activation function.

*4) Support Vector Machine(SVM)*
Support Vector Machines(SVM) aims to find a hyperplane that separates different classes in the data. The hyperplane is the one that has the largest margin between the hyperplane and the closest data point from each class. This hyperplane divides the entire sample space into two regions which corresponds to two classes. Then the instance features can be fed in the model to see which class it falls into. In this experiment, support vector machine with radial basis function kernel is used for training and testing.

*5) Logistic Regression*
Logistic Regression is also a classification algorithm that predicts the probability of one instance belonging to one specific class. Basically, it transforms the result of the linear regression to the probability for the given class using a sigmoid function.

*6) K-Nearest Neighbor*
K-Nearest Neighbor(kNN) is the classical classification algorithm that is intuitive and robust. It assigns the class to the test data instance by considering k nearest neighbor in the training data instances. The final decision is given by the average result given by these k neighbors. In this experiment, the k-Nearest Neighbor uses the Euclidean distance as the proximity measure and k is set to be 5.

### C. Evaluation Metrics

We use the confusion matrix to summarize the correctly and incorrectly classified samples of the testing data using the machine learning algorithms. In addition, for the detailed performance evaluation, we analyze true positive rate (TPR), true negative rate (TNR), false-positive rate (FPR), false-negative rate (FNR), and accuracy for these algorithms. Lastly, we measure the runtime of the training and testing process for each machine learning algorithm. The detailed description of the evaluation metric can be found in Table II and Table III.

TABLE II. CONFUSION MATRIX

| Classified as | Truth | |
|---|---|---|
| | *Phishing* | *Legitimate* |
| Phishing | True Positive(TP) | False Positive(FP) |
| Legitimate | False Negative(FN) | True Negative(TN) |

TABLE III. PERFORMANCE METRIC

| Measures | Description | Formula |
|---|---|---|
| TPR | Number of URLs that are classified as phishing out of total phishing URLs. | TP/(TP+FN) |
| FNR | Number of URLs that are classified as legitimate out of total phishing URLs. | FN/(TP+FN) |
| TNR | Number of URLs that are classified as legitimate out of total legitimate URLs. | TN/(TN+FP) |

| Measures | Description | Formula |
|---|---|---|
| FPR | Number of URLs that are classified as phishing out of total legitimate URLs. | FP/(TN+FP) |
| Precision | Number of true phishing URLs out of URLs classified as phishing | TP/(TP+FP) |
| Recall | Same as TPR | TP/(TP+FN) |
| Accuracy | Number of correctly classifed URLs out all URLs | (TP+TN)/(TP+TN+FP+FN) |

*D. Results*

We apply the machine learning algorithms to the dataset with 10,000 legitimate and 9,965 phishing URLs. The results are tabulated in the Table IV. We then compare the classification results for the same algorithm with and without the keyword feature. For example, row 1(Random Forest) shows the result of algorithm random forest with traditional features only. Row 2(Random Forest(keyword)) shows the result of the algorithm random forest with both the traditional and keyword feature. The one that achieves better result is bolded.

TABLE IV. CLASSIFICATION RESULTS FOR VARIOUS MACHINE LEARNING ALGORITHMS FOR LARGE DATASET

| Algorithms | TPR (%) | FNR (%) | TNR (%) | FPR (%) | Recall (%) | Accuracy (%) |
|---|---|---|---|---|---|---|
| Random Forest | 99.02 | 0.979 | 99.76 | 0.238 | 99.02 | 99.51 |
| Random Forest (keyword) | **99.23** | **0.775** | **99.90** | **0.099** | **99.23** | **99.68** |
| XGboost | 98.08 | 1.917 | 99.70 | 0.297 | 98.08 | 99.17 |
| XGboost (keyword) | **98.32** | **1.672** | **99.74** | **0.258** | **98.32** | **99.28** |
| MLP | 98.16 | 1.835 | 99.24 | 0.753 | 98.16 | 98.89 |
| MLP (keyword) | **98.65** | **1.346** | **99.46** | **0.535** | **98.65** | **99.20** |
| SVM | 97.39 | 2.611 | 99.26 | 0.734 | 97.39 | 98.65 |
| SVM (keyword) | **98.20** | **1.795** | **99.48** | **0.515** | **98.20** | **99.06** |
| Logistic Regression | 93.14 | 6.854 | 97.32 | 2.678 | 93.14 | 95.95 |
| Logistic Regression (keyword) | **94.49** | **5.507** | **98.11** | **1.884** | **94.49** | **96.93** |
| kNN | 97.63 | 2.366 | 99.48 | 0.515 | 97.63 | 98.87 |
| kNN (keyword) | **98.16** | **1.835** | 9950 | **0.496** | **98.16** | **99.06** |

As the classification algorithm can easily achieve more than 95% accuracy in this task. To see the enhancement of our method, we create a subsample of this large dataset by sampling only 10% of it. This small subsample dataset contains 1,000 legitimate and 996 phishing URLs. Similarly, the results are summarized in the Table V.

TABLE V. CLASSIFICATION RESULTS FOR VARIOUS MACHINE LEARNING ALGORITHMS FOR SMALL DATASET

| Algorithms | TPR (%) | FNR (%) | TNR (%) | FPR (%) | Recall (%) | Accuracy (%) |
|---|---|---|---|---|---|---|
| Random Forest | 97.64 | 2.352 | 99.59 | 0.404 | 97.64 | 98.93 |
| Random Forest (keyword) | **97.64** | **2.352** | **99.59** | **0.404** | **97.64** | **98.93** |
| XGboost | 96.86 | 3.137 | 97.57 | 2.429 | 96.86 | 97.32 |
| XGboost (keyword) | **97.65** | **2.352** | **98.38** | **1.619** | **97.65** | **98.13** |
| MLP | 91.76 | 8.235 | 96.96 | 3.036 | 91.76 | 95.19 |
| MLP (keyword) | **94.12** | **5.882** | **98.58** | **1.417** | **94.12** | **97.06** |
| SVM | **94.51** | **5.490** | 97.36 | 2.631 | **94.51** | 96.39 |
| SVM (keyword) | 94.12 | 5.882 | **98.58** | **1.417** | 94.12 | **97.06** |
| Logistic Regression | 90.58 | 9.411 | 96.35 | 3.643 | 90.58 | 94.39 |
| Logistic Regression (keyword) | **92.55** | **7.451** | **97.16** | **2.834** | **92.55** | **95.59** |
| kNN | 94.90 | 5.098 | 97.57 | 2.429 | 94.90 | 96.66 |
| kNN (keyword) | **95.29** | **4.705** | **97.97** | **2.024** | **95.29** | **97.06** |

To analyze the performance for each algorithm, we plot out the graph for the error rate in percentage in the following Fig.2:

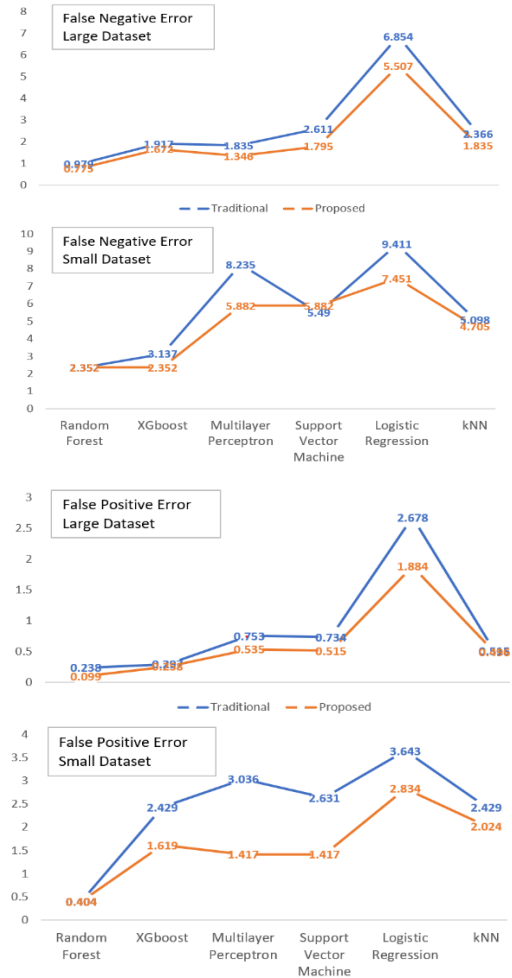

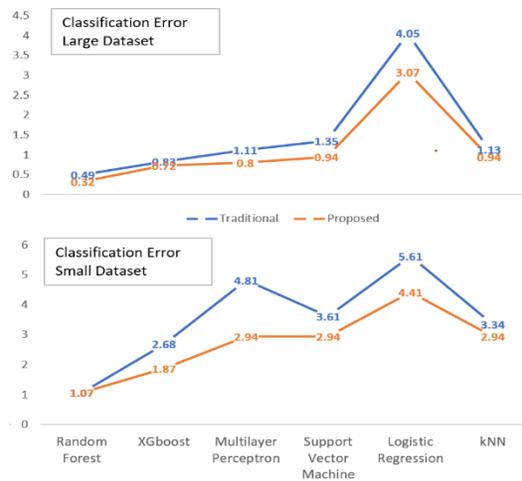

Fig. 2. Errors for different machine learning algorithms on large/small dataset.

*E. Result Interpretation*

We can clearly see that adding the keyword features enhances the classification performance of all the existing machine learning algorithms. Such enhancement is even more significant in the small dataset compared to the large dataset.

Take the multilayer perceptron as an example. It is shown above that the false negative error rate drops from the 1.835% to 1.346% for the large dataset and from 8.235% to 5.882% for the small dataset by incorporating the keyword features. This corresponds to about 30% decrease of the error. As for the false positive error rate, it drops from 0.753% to 0.535% for the large dataset and from 3.036% to 1.417% for the small dataset. This is significant as it decreases the false positive error rate for the small dataset by half.

Specifically, to analyze the contribution of each feature to the classification result, we plot out the feature importance graph for the XGboost algorithm. Basically, this graph measures how much the feature can contribute to the improvement of the performance measure during every tree split. Ten most important features are plotted in the following Fig.3:

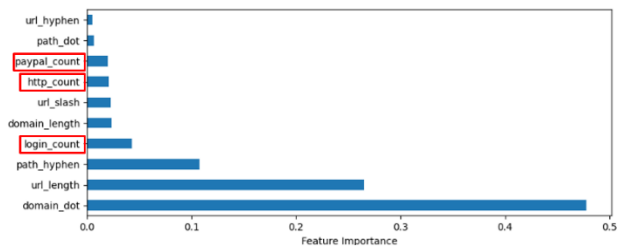

Fig. 3. Feature importance of different keywords for XGboost.

In align with the previous research, several popular features such as the number of dots in the domain, the length of the URL and the length of the domain and so on, are significant in this classification task. The features we proposed are also useful and significant as they are ranked on the top 10 important features. Based on this experiment, we are surprised that the number of the keyword 'login' is the most important keyword. As the attackers are mostly interested in the end-users' account information, the keyword 'login' may be the frequently used word in the malicious websites.

## V. CONCLUSION AND FUTURE WORK

In this paper, a novel hybrid method is developed to enhance the existing machine learning algorithms for phishing URL detection. It combines the traditional features with the keyword features to improve the accuracy. The keyword features are proven to be effective and significant in distinguishing the phishing URLs from the legitimate URLs.

Based on the experimental results on the large dataset and the small dataset, we can observe a notable enhancement on the classification accuracy across various machine learning algorithms. To sum up, this approach has the following advantages: (1)use only the URL information; (2)improve the existing classification algorithms, especially in the small dataset classification tasks; (3)easy to extend to achieve better performance.

Future direction for extending this paper would be automating the process of finding such keywords. More sophisticated and systematic ways could be developed to further exploit the explanatory power of these keywords. Also, the underlying meaning of the keywords can be further discussed.